%% file: linear-sampling.tex
\documentclass[11pt]{article}
\usepackage{fullpage}
\usepackage{latexsym}
\usepackage{mathptmx}
\usepackage[scaled]{helvet}
\usepackage[dvips]{graphicx}
\usepackage{amsmath, amssymb, amsthm, amsfonts}

\newcommand{\eat}[1]{}

\newcommand{\pr}{\mathbb{P}}
\newcommand{\ex}{\mathbb{E}}

\newtheorem{fact}{Fact}

\newtheorem{lemma}{Lemma}
\newtheorem{theorem}{Theorem}

\newtheorem{definition}{Definition}

\title{A Linear-time Algorithm for Sparsification of Unweighted Graphs}
\author{Ramesh Hariharan\\Strand Life Sciences \and
Debmalya Panigrahi\\CSAIL, MIT}
\date{}

\begin{document}

\maketitle

\begin{abstract}
Given an undirected graph $G$ and an error parameter $\epsilon > 0$, 
the {\em graph sparsification} problem requires
sampling edges in $G$ and giving the sampled 
edges appropriate weights to obtain a sparse graph $G_{\epsilon}$ with the following
property: the weight of every cut in $G_{\epsilon}$ is within a factor of
$(1\pm \epsilon)$ of the weight of the corresponding cut in $G$. If $G$ is
unweighted, an $O(m\log n)$-time algorithm for constructing $G_{\epsilon}$ with 
$O(n\log n/\epsilon^2)$ edges in expectation, and an $O(m)$-time algorithm for
constructing $G_{\epsilon}$ with $O(n\log^2 n/\epsilon^2)$ edges in expectation 
have recently been developed~\cite{HariharanP10}. In this paper, we improve these 
results by giving an $O(m)$-time algorithm 
for constructing $G_{\epsilon}$ with $O(n\log n/\epsilon^2)$
edges in expectation, for unweighted graphs. Our algorithm is optimal in terms
of its time complexity; further, no efficient algorithm is known for constructing
a sparser $G_{\epsilon}$. Our algorithm is Monte-Carlo, i.e. it produces the 
correct output with high probability, as are all efficient graph sparsification
algorithms.
\end{abstract}

\newpage

\input introduction

\input framework-outline

\input algorithm

\input future

\bibliographystyle{plain}
\bibliography{ref}

\appendix

\input picon

\end{document}

%% file: introduction.tex
\section{Introduction}
\label{sec:intro}

A {\em cut} of an undirected graph is a partition of its
vertices into two disjoint sets. The {\em weight} of a
cut is the sum of weights of the edges crossing the cut, 
i.e. edges having one endpoint each in the two vertex 
subsets of the partition. For unweighted graphs, each edge
is assumed to have unit weight. Cuts play an important
role in many problems in graphs: e.g., the 
maximum flow between a pair of vertices is equal to the
minimum weight cut separating them.

A {\em skeleton} $G'$ of an undirected graph $G$ is a subgraph 
of $G$ on the same set of vertices where each edge in $G'$
can have an arbitrary weight. The problem of finding an appropriately
weighted sparse skeleton for an undirected graph $G$ that approximately
preserves the weights of all cuts in $G$ was introduced and studied by
Karger {\em et al} in a series of 
results~\cite{Karger94a, Karger94b, BenczurK96} culminating in the
following theorem. Throughout this paper, 
for any undirected graph $G$ and any $\epsilon\in (0, 1]$, 
$(1\pm \epsilon)G$ will denote the set of all appropriately weighted 
subgraphs of $G$ where the weight of every cut in the subgraph
is within a factor of $(1\pm \epsilon)$ of the weight of the 
corresponding cut in $G$.
\begin{theorem}[Bencz\'ur-Karger~\cite{BenczurK96}]
\label{thm:sampling}
For any undirected graph $G$ with $m$ edges and
$n$ vertices, and for any error parameter $\epsilon\in (0, 1]$, 
a skeleton $G_{\epsilon}$ containing 
$O(\frac{n\log n}{\epsilon^2})$ edges in expectation such that 
$G_{\epsilon}\in (1\pm\epsilon)G$ 
with high probability\footnote{We say that a property holds
{\em with high probability} (or whp) for a graph on $n$ vertices
if its failure probability can be bounded by the inverse of
a fixed polynomial in $n$.} can be found in $O(m\log^2 n)$ time
if $G$ is unweighted and $O(m\log^3 n)$ time otherwise.
\end{theorem}
\noindent 
Besides its combinatorial ramifications, the importance of this 
result stems from its use as a pre-processing step in several graph 
algorithms, 
e.g. to obtain an $\tilde{O}(n^{3/2}+m)$-time algorithm for approximate 
maximum flow using the $\tilde{O}(m^{3/2})$-time algorithm for exact maxflow 
due to Goldberg and Rao~\cite{GoldbergR98}; and more recently, 
$\tilde{O}(n^{3/2}+m)$-time algorithms for approximate sparsest 
cut~\cite{KhandekarRV09, Sherman09}. 

Subsequent to Bencz\'ur and Karger's work,
Spielman and Teng~\cite{SpielmanT04, SpielmanT06} extended their results to 
preserving all quadratic forms, of which cuts are a special case;
however, the size of the skeleton constructed was $O(n\log^c n)$
for some large constant $c$. Spielman and 
Srivastava~\cite{SpielmanS08} improved this result by constructing skeletons of
size $O(\frac{n\log n}{\epsilon^2})$ in $O(m\log^{O(1)} n)$ time, 
while continuing to preserve all quadratic 
forms. Recently, this result was further improved by Batson {\em et al}~\cite{BatsonSS09} 
who gave a deterministic algorithm for constructing skeletons of 
size $O(\frac{n}{\epsilon^2})$ that preserve the weights of all cuts
whp. While their result is optimal in terms of
the size of the skeleton constructed, the time complexity of their
algorithm is $O(\frac{mn^3}{\epsilon^2})$, rendering it somewhat useless 
in terms of applications. 

Recently, further progress has been made on efficiently constructing a
skeleton graph in the form of the following theorem due to Hariharan and
Panigrahi~\cite{HariharanP10}.
\begin{theorem}[Hariharan-Panigrahi~\cite{HariharanP10}]
\label{thm:sampling-hp}
For an undirected graph $G$ with $m$ edges and
$n$ vertices, and for any error parameter $\epsilon\in (0, 1]$,
the following algorithmic results can be obtained for constructing a skeleton 
graph $G_{\epsilon}$ that is in $(1\pm\epsilon)G$ with high probability:
\begin{itemize}
\item If the expected number of edges in $G_{\epsilon}$ is 
$O(n\log^2 n/\epsilon^2)$, then $G_{\epsilon}$ can be constructed in
$O(m)$ time if $G$ has polynomial edge weights and $O(m\log^2 n)$
time if $G$ has arbitrary edge weights.
\item If the expected number of edges in $G_{\epsilon}$ is 
$O(n\log n/\epsilon^2)$, then $G_{\epsilon}$ can be constructed in
$O(m\log n)$ time if $G$ is unweighted, and
$O(m\log^2 n)$ time if $G$ has polynomial edge weights.
\end{itemize}
\end{theorem}
\noindent 
Combining the above two results, one can obtain an algorithm to
construct a skeleton graph $G_{\epsilon}$ that preserves the 
weights of all cuts whp and has $O(n\log n/\epsilon^2)$ edges in
expectation in 
$O(m + n\log^4 n/\epsilon^2)$ time if $G$ has polynomial edge weights.

A natural conclusion for this line of work would be to obtain an
$O(m)$-time algorithm for constructing a skeleton graph $G_{\epsilon}$
containing $O(n\log n/\epsilon^2)$ edges in expectation. In this
correspondence, we obtain this result in the form of the following
theorem if $G$ is unweighted. It may
be noted that even if $G$ is unweighted, the best sparsification 
result known previously was Theorem~\ref{thm:sampling-hp}.
\begin{theorem}
\label{thm:sampling-unweighted}
For an undirected unweighted graph $G$ with $m$ edges and
$n$ vertices, and for any error parameter $\epsilon\in (0, 1]$,
a skeleton graph $G_{\epsilon}$ that is in $(1\pm\epsilon)G$ 
with high probability and has $O(n\log^2 n/\epsilon^2)$ edges
in expectation, can be constructed in $O(m)$ time.
\end{theorem}
\noindent
Note that the above algorithm is optimal in terms of its running
time, and there is no efficient algorithm known for constructing a 
sparser skeleton, even for unweighted graphs. (As mentioned previously,
the only algorithm known that constructs a sparser skeleton has a 
time complexity of $O(n^3m/\epsilon^2)$~\cite{BatsonSS09}.)

Before describing our algorithm in more detail, it is worth mentioning
some of the related research in graph sparsification. In a recent result,
Fung and Harvey~\cite{FungH10} show that sampling uniformly random 
spanning trees of a graph produces good sparsifiers. This approach
was used previously to obtain coarser sparsifiers by 
Goyal {\em et al}~\cite{GoyalRV09}. Fung and Harvey also show
that sampling edges according to their standard connectivities also 
produces good sparsifiers, a result obtained independently by Hariharan
and Panigrahi~\cite{HariharanP10}. The problem of
graph sparsification in the semi-streaming model was first considered
by Ahn and Guha~\cite{AhnG09} who gave a one-pass algorithm for constructing a
skeleton $G_{\epsilon}$ containing $O(n\log n\log (m/n)/\epsilon^2)$ edges. 
Recently, Goel {\em et al} have given the following algorithms for this 
problem~\cite{GoelKK10}:
\begin{itemize}
\item An $O(m\log\log n)$-time one-pass algorithm for constructing a 
skeleton graph $G_{\epsilon}$ containing $O(n\log^2 n/\epsilon^2)$ edges
in expectation. The size of the skeleton can be improved to 
$O(n\log n/\epsilon^2)$ edges; however, the time complexity of the
algorithm then becomes $O(m\log\log n + n\log^5 n/\epsilon^2)$.
\item An $O(m)$-time two-pass algorithm for constructing a skeleton graph 
$G_{\epsilon}$ containing $O(n\log n/\epsilon^2)$ edges
in expectation, if $m = \Omega(n^{1+\delta})$ for some constant $\delta > 0$.
\end{itemize}
Observe that both results, if applied to a non-streaming model, are weaker
than Theorem~\ref{thm:sampling-hp}. Another area of recent interest, though
not directly related to our problem, is that of vertex 
sparsification~\cite{Moitra09, LeightonM10}. Given a graph $G = (V, E)$ and
a subset of vertices $S\subset V$, the goal here is to create a graph 
$G_S = (S, E_S)$ that approximately preserves some desired connectivity
property of $G$ (e.g. minimum steiner cut~\cite{Moitra09}, maximum 
multi-commodity flow~\cite{LeightonM10}).

\subsection{Our Techniques}
All previous algorithms for graph sparsification have two phases: in the first
phase, a suitable probability $p_e$ for sampling each edge $e$ is determined; and
then, in the second phase, every edge is independently sampled with probability
$p_e$ and given weight $1/p_e$ in the skeleton graph if selected in the sample. 
Our main technical novelty is in interleaving the sampling process with that of 
estimating sampling probabilities. Such interleaving leads to several technical 
hurdles:
\begin{itemize}
\item It introduces dependence between the sampling of different edges. Such
dependence has appeared previously in sparsification algorithms for the 
semi-streaming model, but the nature of the dependence in our algorithm is
somewhat different from that in the streaming algorithms.
\item An edge may now be sampled multiple times, and errors are accrued 
in each such sampling. This requires us to choose the interleaved sampling 
probabilities very carefully so that the errors do not add up.
\end{itemize}

At a high level, our algorithm has the same iterative structure as algorithms
in~\cite{BenczurK96} and \cite{HariharanP10}. In each iteration, the algorithm 
identifies suitable sampling probabilities of a subset of edges and removes them 
from the graph. It is in what the algorithm does with the remaining edges that
our algorithm differs from previous work. While all remaining edges are retained
for the next iteration in previous algorithms, we sample all the edges with 
probability 1/2 and retain only half of them in expectation for the next iteration.
The intuition behind this sampling comes from the observation that the sampling
probabilities decrease (approximately by a factor of 2) with every iteration; 
therefore, a natural approach is to sample the remaining edges with probability 
1/2 and retain only the selected edges thereby reducing the time complexity of
the next iteration.

Suppose $X_i$ be the set of remaining edges at the beginning of iteration $i$,
$F_i$ be the set of edges whose sampling probabilities are determined in 
iteration $i$ and $Y_i = X_i\setminus F_i$ be the set of remaining edges
after iteration $i$. (Note that $X_{i+1}$ is therefore constructed by sampling 
each edge in $Y_i$ with probability 1/2.)
Our proof technique consists of two parts. First, we show that the graph $S$ 
containing appropriately weighted edges in $\cup_i F_i$
is in $(1\pm \epsilon/3)G$ whp, i.e. even though edges in
$Y_i\setminus X_{i+1}$ are sampled out between iterations $i$ and $i+1$ for each
$i$, the retained edges (when weighted appropriately) are sufficient to preserve 
all cuts. In the second step of the proof, we show that the skeleton graph 
$G_{\epsilon}$ constructed by sampling edges in $\cup_i F_i$ and giving them
appropriate weights is in $(1\pm \epsilon/3)S$ whp. For this proof, we use the
generic sparsification framework developed recently by Hariharan and 
Panigrahi~\cite{HariharanP10}. Combining these two steps, we conclude that 
$G_{\epsilon}\in (1\pm\epsilon)G$ whp.

\noindent
\paragraph{Roadmap.} In section~\ref{sec:framework}, we give an 
outline of the generic sampling framework from~\cite{HariharanP10}
that we use later in our proof. In section~\ref{sec:algorithm}, we
describe our sparsification algorithm, prove its correctness and
derive its time complexity. Finally, we conclude with some open
questions in section~\ref{sec:future}.

%% file: framework-outline.tex
\section{Preliminaries}
\label{sec:framework}

We first need to introduce the notion of $k$-heavy edges,
for any $k > 0$. 
\begin{definition}
An edge $e = (u, v)$ of an undirected graph $G = (V, E)$ is said to 
be {\em $k$-heavy} if the maximum flow between vertices $u$ and $v$ in $G$
is at least $k$. 
\end{definition}
By Menger's theorem (see e.g.,~\cite{CormenLRS01}), it follows that if
$e= (u, v)$ is $k$-heavy, then the weight of every cut in $G$ having 
$u$ and $v$ on different sides is at least $k$. 

\subsection{Outline of Sparsification Framework from~\cite{HariharanP10}}
Suppose $G = (V, E)$ is an undirected graph where edge $e\in E$ 
has a positive integer weight $w_e$. 
Let $G_M = (V, E_M)$ denote the multi-graph
constructed by replacing each edge $e$ by 
$w_e$ unweighted parallel edges $e_1, e_2, \ldots, e_{w_e}$. 
Consider any $\epsilon\in (0, 1]$.
Suppose we construct a skeleton $G_{\epsilon}$ where each edge 
$e_{\ell}\in E_M$ is present in graph $G_{\epsilon}$ independently
with probability $p_e$, and if present, it is given a weight of 
$1/p_e$.
Let $p_e = \min(\frac{96\alpha\ln n}{0.38\lambda_e\epsilon^2}, 1)$, 
where $\alpha$ is independent of $e$ and $\lambda_e$ is some 
parameter of $e$ satisfying $\lambda_e\leq 2^n-1$.  
The authors describe a sufficient condition that characterizes
a good choice of $\alpha$ and $\lambda_e$'s.

To describe this sufficient condition, partition the edges 
in $G_M$ according to the value of 
$\lambda_e$ into sets $R_0, R_1, \ldots, R_K$ where 
$K = \lfloor\lg \max_{e\in E}\{\lambda_e\}\rfloor\leq n-1$ and
$e_i\in R_j$ iff $2^j\leq \lambda_e\leq 2^{j+1}-1$.
Now, let ${\bf Q} = (Q_0, Q_1, Q_2, \ldots, Q_i = (V, W_i), \ldots, Q_k)$ 
be a sequence of subgraphs of $G_M$ (edges of $G_M$ are allowed to be replicated
multiple times in the $Q_i$s) such that $R_i\subseteq W_i$ for every $i$. 
$\bf Q$ is said to be a {\em $(\pi, \alpha)$-certificate} corresponding to
the above choice of $\alpha$ and $\lambda_e$'s if the following properties are satisfied:
\begin{description}
\item[$\pi$-connectivity] 
For $i\geq 0$, any edge $e_{\ell}\in R_i$ is $\pi$-heavy in $Q_i$.
\item[$\alpha$-overlap] For any cut $C$ containing $c$ edges in $G_M$, 
let $w^{(C)}_i$ be the number
of edges that cross $C$ in $Q_i$. Then, for all cuts $C$,
$\sum_{i=0}^k \frac{w^{(C)}_i 2^{i-1}}{\pi}\leq \alpha c$.
\end{description}
Then, the following theorem holds.
 
\begin{theorem}[Hariharan-Panigrahi~\cite{HariharanP10} (Theorem 8)]
\label{thm:framework}
If there exists a $(\pi, \alpha)$-certificate for a particular choice of $\alpha$ and $\lambda_e$'s , then 
the skeleton $G_{\epsilon}\in (1\pm\epsilon)G$ with probability at least $1 - 4/n$.
Further $G_{\epsilon}$ has $O(\frac{\alpha \log n}{\epsilon^2}\sum_{e\in E} \frac{w_e}{\lambda_e})$ edges in expectation.
\end{theorem}
\noindent
We also need the following lemma, which is a slight variation 
of Lemma~5 from~\cite{HariharanP10}. (For completeness, we
give a proof in the appendix.) For
an undirected unweighted graph $G = (V, E)$, let $R\subseteq E$
and $Q\supseteq R$ be subsets of edges such that $R$ is 
$\pi$-heavy in $(V, Q)$. Suppose each edge $e\in R$ is sampled
with probability $p$,
and if selected, given a weight of $1/p$ to form a set of weighted
edges $\widehat{R}$. Now, for any cut
$C$ in $G$, let $R^{(C)}$,  $Q^{(C)}$ and $\widehat{R^{(C)}}$ be 
the sets of edges crossing cut $C$ in $R$, $Q$ and 
$\widehat{R}$ respectively;
also let the total weight of edges in $R^{(C)}$, $Q^{(C)}$
and $\widehat{R^{(C)}}$
be $r^{(C)}$, $q^{(C)}$ and $\widehat{r^{(C)}}$ respectively. 
Then the following lemma holds.
\begin{lemma}
\label{lma:picon-hp}
For any $\delta\in (0, 1]$ satisfying 
$\delta^2\cdot p\cdot \pi \geq \frac{6\ln n}{0.38}$,
\begin{equation*}
|r^{(C)}-\widehat{r^{(C)}}| \leq \delta q^{(C)}
\end{equation*}
for all cuts $C$ in $G$ with probability at least $1 - 4/n^2$.
\end{lemma}
\subsection{Nagamochi-Ibaraki Forests}
We first introduce the notion of {\em spanning forests} of a graph. 
As earlier, $G$ denotes a graph with 
integer edge weights $w_e$ for edge $e$ and $G_M$ is the 
unweighted multi-graph where $e$ is replaced with $w_e$ parallel
unweighted edges.
\begin{definition}
A {\em spanning forest} $T$ of $G_M$ (or equivalently of $G$) 
is an (unweighted) acyclic subgraph of $G$ satisfying the property 
that any two vertices are connected in $T$ if and only if they are connected in $G$.
\end{definition}
We partition the set of edges in $G_M$ into a set of 
forests $T_1, T_2, \ldots$ using the following rule: 
{\em $T_i$ is a spanning forest of the graph formed by 
removing all edges in $T_1, T_2, \ldots, T_{i-1}$ from $G_M$
such that for any edge $e\in G$, all its copies in $G_M$
appear in a set of contiguous forests 
$T_{i_e}, T_{i_e+1},\ldots, T_{i_e + w_e-1}$}.
This partitioning technique was introduced by Nagamochi and Ibaraki 
in~\cite{NagamochiI92a}, and these forests are known as 
{\em Nagamochi-Ibaraki forests} (or NI forests). The following is
a basic property of NI forests.
\begin{lemma}[Nagamochi-Ibaraki~\cite{NagamochiI92a, NagamochiI92b}]
\label{lma:ni-con}
For any pair of vertices $u, v$, they are connected in 
NI forests $T_1, T_2, \ldots, T_{k(u, v)}$ for some 
$k(u, v)$ and not connected in any forest $T_j$, for
$j > k(u, v)$.
\end{lemma}
\noindent
Nagamochi and Ibaraki also gave an algorithm for constructing 
NI forests that runs in $O(m+n)$ time if $G_M $ is a simple
graph (i.e. $G$ is unweighted) and 
$O(m+n\log n)$ time otherwise~\cite{NagamochiI92a, NagamochiI92b}. 

%% file: algorithm.tex
\section{The Algorithm}
\label{sec:algorithm}

We describe out sparsification algorithm for an unweighted graph 
$G = (V, E)$ with $m$ edges and $n$ vertices, and an error
parameter $\epsilon\in (0, 1]$ as inputs. We prove that the skeleton
graph $G_{\epsilon}$ produced by the algorithm is in $(1\pm\epsilon)G$ 
with high probability. We then show that the expected number of edges in
$G_{\epsilon}$ is $O(n\log n/\epsilon^2)$. Finally, we prove that the
expected time complexity of the algorithm is $O(m)$.

\noindent\paragraph{Description of the Algorithm.}
The algorithm has three phases. The first phase has the following steps:
\begin{itemize} 
\item If $m \leq 2\rho n$, where $\rho = \frac{1014\ln n}{0.38\epsilon^2}$,
$G$ is sparse enough itself. Therefore, we take $G$ as our skeleton graph.
\item Otherwise, we construct a set of NI forests of $G$ and all edges in 
the first $2\rho$ NI forests are included in the skeleton graph
$G_{\epsilon}$ with weight 1. We call these edges $F_0$. 
The edge set $Y_0$ is then defined as $E\setminus F_0$.
\end{itemize}
The second phase is iterative. The input to iteration $i$ is a graph
$(V, Y_{i-1})$, which is a subgraph of the input graph to iteration $i-1$
(i.e. $Y_{i-1}\subseteq Y_{i-2}$). Iteration $i$ comprises the following steps: 
\begin{itemize}
\item If the number of edges in $Y_{i-1}$ is at most $2\rho n$, we take
all those edges in $G_{\epsilon}$ with weight $2^{i-1}$ each, and terminate the
algorithm.
\item Otherwise, all edges in $Y_i$ are sampled with probability 
$1/2$; call the sample $X_i$ and let $G_i = (V, X_i)$.
\item We identify a set of edges in $X_i$ (call this set $F_i$) that has
the following properties:
	\begin{itemize}
	\item The number of edges in $F_i$ is at most $2 k_i |V_c|$, where 
	$k_i = \rho\cdot 2^{i+1}$, and 
	$V_c$ is the set of components in $(V, Y_i)$, where $Y_i = X_i\setminus F_i$.
	\item Each edge in $Y_i$ is $k_i$-heavy in $G_i$.
	\end{itemize}
\item We give a sampling probability $p_i = \min(\frac{3}{169\cdot 2^{2i-9}}, 1)$ to all
edges in $F_i$.
\end{itemize}
The final phase consists of replacing each edge in $F_i$ (for each $i$)
with $2^i$ parallel edges, and then sampling each parallel edge 
independently with probability $p_i$. If an edge is selected in the 
sample, it is added to the skeleton graph $G_{\epsilon}$ with weight
$1/p_i$.

We now give a short description of the sub-routine that constructs 
the set $F_i$ in iteration $i$ of the second phase of the algorithm.
This sub-routine is iterative itself: we start with $V_c = V$ 
and $E_c = X_i$, and let $G_c = (V_c, E_c)$. 
We repeatedly construct $k_i + 1$ NI forests for $G_c$ where 
$k_i = \rho 2^{i+1}+1$ and contract all edges 
in the $(k_i+1)$st forest to obtain
a new $G_c$, until $|E_c|\leq \rho 2k_i |V_c|$. The set of edges $E_c$
that finally achieves this property forms $F_i$.

The complete algorithm is given in Figure~\ref{fig:algo}.
\begin{figure}
\centering
\begin{enumerate}
\item Set $\rho = \frac{1014\ln n}{0.38\epsilon^2}$.
\item If $m \leq 2\rho n$, then $G_{\epsilon} = G$; else, go to step~\ref{start}.
\item\label{start} Construct NI forests $T_1, T_2, \ldots$ for $G$.
\item Set $i = 0$.
\item Set $X_i = E$; $F_i = \cup_{1\leq j\leq 2\rho} T_j$; $Y_i = X_i\setminus F_i$.
\item\label{add1} Add each edge in $F_i$ to $G_{\epsilon}$ with weight 1.
\item\label{loop1} If $|Y_i|\leq 2\rho n$, then add each edge in $Y_i$ to 
$G_{\epsilon}$ with weight $2^{i-1}$ and terminate; 
else, go to step~\ref{loop1-next}. 
\item\label{loop1-next} Sample each edge in $Y_i$ with probability 1/2 to 
construct $X_{i+1}$.
\item Increment $i$ by 1.
\item Set $E_c = X_i$; $V_c = V$.
\item Set $k_i = \rho\cdot 2^{i+1}$.
\item\label{loop2} If $|E_c| \leq 2 k_i |V_c|$, then
\begin{enumerate}
	\item Set $F_i = E_c$; $Y_i = X_i\setminus E_c$.
	\item\label{add2} For each edge $e\in F_i$, set 
	$\lambda_e = \rho\cdot 4^i$.
	\item Go to step~\ref{loop1}.
\end{enumerate}
Else,
\begin{enumerate}
	\item Construct NI forests $T_1, T_2, \ldots, T_{k_i + 1}$ for graph 
	$G_c = (V_c, E_c)$. 
	\item Update $G_c = (V_c, E_c)$ by contracting all edges in $T_{k_i + 1}$.
  \item Go to step~\ref{loop2}.
\end{enumerate}
\item For each edge $e\in \cup_i F_i$,
	\begin{enumerate}
	\item Set $p_e = \min(\frac{9216\ln n}{0.38 \lambda_e \epsilon^2}, 1) = \min(\frac{3}{169\cdot 2^{2i-9}}, 1)$.
	\item Generate $r_e$ from {\bf Binomial}$(2^i, p_e)$.
	\item If $r_e > 0$, add edge $e$ to $G_{\epsilon}$ with weight $r_e/p_e$.
	\end{enumerate}
\end{enumerate}
\caption{Our sparsification algorithm}
\label{fig:algo}
\end{figure}

\noindent\paragraph{Cut Preservation.}
We first show that the skeleton graph $G_{\epsilon}$
produced by the above algorithm is in $(1\pm\epsilon)G$ with high probability. 
We use the following notation throughout: 
for any set of unweighted edges $Z$, $cZ$ denotes these edges with
a weight of $c$ given to each edge. 

Our goal is to prove the following theorem.
\begin{theorem}
\label{thm:final}
$G_{\epsilon}\in (1\pm \epsilon)G$ with probability at least $1-8/n$.
\end{theorem}
\noindent
As outlined in the introduction, our proof has two stages. Let 
$K$ be the maximum value of $i$ for which $F_i\not= \emptyset$;
let $S = \left(\cup_{i=0}^K 2^i F_i\right)\cup 2^K Y_K$ and 
$G_S = (V, S)$. Then, in the first stage, 
we prove the following theorem.
\begin{theorem}
\label{thm:stage1}
$G_S\in (1\pm \epsilon/3)G$ with probability at least $1-4/n$.
\end{theorem}
\noindent 
In the second stage, we prove the following theorem.
\begin{theorem}
\label{thm:stage2}
$G_{\epsilon}\in (1\pm \epsilon/3)G_S$ with probability at least $1-4/n$.
\end{theorem}
\noindent
Combining the above two theorems and using the union bound, we obtain
Theorem~\ref{thm:final}. (Observe that since $\epsilon\leq 1$, 
$(1 + \epsilon/3)^2\leq 1 + \epsilon$ and 
$(1 - \epsilon/3)^2\geq 1 - \epsilon$).

The following property is key to proving both Theorem~\ref{thm:stage1}
and Theorem~\ref{thm:stage2}.
\begin{lemma}
\label{lma:heavy}
For any $i\geq 0$, any edge $e\in Y_i$ is $k_i$-heavy in 
$G_i = (V, X_i)$, where $k_i = \rho\cdot 2^{i+1}$.
\end{lemma}
\begin{proof}
For $i=0$, all edges in $Y_i$ are in NI forests 
$T_{2\rho+1}, T_{2\rho+2}, \ldots$ of $G_i = G$. 
The proof follows from Lemma~\ref{lma:ni-con}.

We now prove the lemma for $i\geq 1$.
Let $G_e = (V_e, E_e)$ be the component of $G_i$ containing 
$e$. We will show that $e$ is $k_i$-heavy in 
$G_e$; since $G_e$ is a subgraph of $G_i$, the lemma follows. 
In the execution of the else block of step~\ref{loop2} on
$G_e$, there are multiple contraction operations, each of 
them comprising the contraction of a set of edges. We show
that any such contracted edge is $k_i$-heavy
in $G_e$; it follows that $e$ is $k_i$-heavy in $G_e$. 
 
Let $G_e$ have $t$ contraction phases and let the graph
produced after contraction phase $r$ be $G_{e, r}$. 
We now prove that all edges contracted in phase $r$
must be $k_i$-heavy in $G_e$ by induction 
on $r$. For $r=1$, since $e$ appears in the
$(k_i+1)$st NI forest of phase 1, $e$ is 
$k_i$-heavy in $G_e$ by Lemma~\ref{lma:ni-con}. 
For the inductive
step, assume that the property holds for phases
$1, 2, \ldots, r$. Any edge that is contracted in 
phase $r+1$ appears in the $(k_i + 1)$st NI forest 
of phase $r+1$; therefore, $e$ is $k_i$-connected 
in $G_{e, r}$ by Lemma~\ref{lma:ni-con}. 
By the inductive hypothesis, 
all edges of $G_e$ contracted 
in previous phases are $k_i$-heavy in $G_e$; 
therefore, an edge that is $k_i$-heavy in $G_{e, r}$
must have been $k_i$-heavy in $G_e$.
\end{proof}

\noindent
\paragraph{Proof of Theorem~\ref{thm:stage1}.}
The next lemma follows from Lemma~\ref{lma:picon-hp}.
\begin{lemma}
\label{lma:step}
With probability at least $1-4/n^2$, for every cut $C$ in $G_i$, 
$|2x^{(C)}_{i+1} + f^{(C)}_i - x^{(C)}_i| \leq \frac{\epsilon/13}{2^{i/2}}\cdot x^{(C)}_i$.
\end{lemma}
\begin{proof}
Use the following parameters in Lemma~\ref{lma:picon-hp}:
\begin{itemize} 
\item $R = Y_i$; $Q = X_i$; $\widehat{R} = 2X_{i+1}$ 
\item $\delta = \frac{\epsilon/13}{2^{i/2}}$; $p = 1/2$; $\pi = \rho\cdot 2^{i+1}$.
\end{itemize}
\noindent
Lemma~\ref{lma:heavy} ensures that $R$ is $\pi$-heavy in $(V, Q)$; 
also, it can be verified that $\delta^2\cdot p\cdot \pi = 6\ln n$.
\end{proof}
\noindent
We use the above lemma to prove the following lemma. 
\begin{lemma}
\label{lma:sampling1}
Let $S_j = \left(\cup_{i=j}^K 2^{i-j} F_i\right)\cup 2^{K-j}Y_K$ 
for any $j\geq 0$. Then,
$S_j \in (1\pm(\epsilon/3)2^{-j/2})G_j$ with probability at least 
$1-4/n$, where $G_j = (V, X_j)$.
\end{lemma}
\noindent
To prove this lemma, we need to use the following fact.
\begin{fact}
\label{fact:product}
Let $x\in (0, 1]$ and $r_i = 13\cdot 2^{i/2}$. Then,
for any $k\geq 0$,
\begin{eqnarray*}
\prod_{i = 0}^k (1 + x/r_i) & \leq & 1 + x/3 \\
\prod_{i = 0}^k (1 - x/r_i) & \geq & 1 - x/3.
\end{eqnarray*}
\end{fact}
\begin{proof}
We prove by induction on $k$. For $k=0$, the property trivially holds.
Suppose the property holds for $k-1$. Then,
\begin{eqnarray*}
\prod_{i = 0}^k (1 + x/r_i)
& = & \prod_{i = 0}^k (1 + \frac{x}{13\cdot 2^{i/2}}) \\
& = & (1 + x/13)\cdot \prod_{i = 1}^k \left(1 + \frac{x/\sqrt{2}}{13\cdot 2^{(i-1)/2}}\right) \\
& \leq & (1 + x/13)\cdot (1 + x/(3\sqrt{2})) \\
& \leq & 1 + x/3 \\
\prod_{i = 0}^k (1 - x/r_i)
& = & \prod_{i = 0}^k (1 - \frac{x}{13\cdot 2^{i/2}}) \\
& = & (1 - x/13)\cdot \prod_{i = 1}^k \left(1 - \frac{x/\sqrt{2}}{13\cdot 2^{(i-1)/2}}\right) \\
& \geq & (1 - x/13)\cdot (1 - x/(3\sqrt{2})) \\
& \geq & 1 - x/3.
\end{eqnarray*}
\end{proof}
\noindent
\begin{proof}[Proof of Lemma~\ref{lma:sampling1}]
For any cut $C$ in $G$, let the edges crossing $C$ in $S_j$ be 
$S^{(C)}_j$, and let their total weight be $s^{(C)}_j$. Also,
let $X^{(C)}_i$, $Y^{(C)}_i$ and $F^{(C)}_i$ be the set of edges 
crossing cut $C$ in $X_i$, $Y_i$ and $F_i$ respectively,
and let their total weights be $x^{(C)}_i$, $y^{(C)}_i$ and 
$f^{(C)}_i$. (Recall that all edges in $X_i$, $Y_i$ and $F_i$ 
are unweighted; therefore $x^{(C)}_i = |X^{(C)}_i|$,
$y^{(C)}_i = |Y^{(C)}_i|$ and $f^{(C)}_i = |F^{(C)}_i|$.)

Since $K\leq n-1$, we can use the union bound on Lemma~\ref{lma:step}
to conclude
that with probability at least $1-4/n$, for every $0\leq i\leq K$
and for all cuts $C$, 
\begin{eqnarray*}
2x^{(C)}_{i+1}+f^{(C)}_i & \leq & (1+\epsilon/r_i) x^{(C)}_i \\
2x^{(C)}_{i+1}+f^{(C)}_i & \geq & (1-\epsilon/r_i) x^{(C)}_i,
\end{eqnarray*}
where $r_i = 13\cdot 2^{i/2}$. Then,
\begin{eqnarray*}
   s^{C}_j 
  & = & 2^{K-j} y^{(C)}_K + 2^{K-j} f^{(C)}_K + 2^{K-1-j} f^{(C)}_{K-1} + \ldots + f^{(C)}_j \\
  & = & 2^{K-j} x^{(C)}_K + 2^{K-1-j} f^{(C)}_{K-1} + \ldots + f^{(C)}_j\quad{\rm since~} y^{(C)}_K + f^{(C)}_K = x^{(C)}_K \\
    & = & 2^{K-1-j}(2x^{(C)}_K + f^{(C)}_{K-1}) + (2^{K-2-j} f^{(C)}_{K-2}+ \ldots + f^{(C)}_j) \\
  & \leq & (1+\epsilon/r_{K-1}) 2^{K-1-j} x^{(C)}_{K-1} + (2^{K-2-j} f^{(C)}_{K-2}+ \ldots + f^{(C)}_j) \\
  & \leq & (1+\epsilon/r_{K-1}) (2^{K-1-j} x^{(C)}_{K-1} + 2^{K-2-j} f^{(C)}_{K-2}+ \ldots + f^{(C)}_j) \\
  & \ldots & \\
  & \leq & (1+\epsilon/r_{K-1}) (1+\epsilon/r_{K-2}) \ldots (1+\epsilon/r_j) x^{(C)}_j \\
  & \leq & (1+(\epsilon 2^{-j/2})/r_{K-1-j}) (1+(\epsilon 2^{-j/2})/r_{K-2-j}) \ldots (1+(\epsilon 2^{-j/2})/r_0) x^{(C)}_j\quad{\rm since~} r_{j+i} = r_i \cdot 2^{j/2}\\
  & \leq & (1+(\epsilon/3)2^{-j/2}) x^{(C)}_j\quad \rm{by~Fact~\ref{fact:product}.}
\end{eqnarray*}
Similarly,
\begin{eqnarray*}
   s^{C}_j 
  & = & 2^{K-j} y^{(C)}_K + 2^{K-j} f^{(C)}_K + 2^{K-1-j} f^{(C)}_{K-1} + \ldots + f^{(C)}_j \\
  & = & 2^{K-j} x^{(C)}_K + 2^{K-1-j} f^{(C)}_{K-1} + \ldots + f^{(C)}_j\quad{\rm since~} y^{(C)}_K + f^{(C)}_K = x^{(C)}_K\\
  & = & 2^{K-1-j}(2x^{(C)}_K + f^{(C)}_{K-1}) + (2^{K-2-j} f^{(C)}_{K-2}+ \ldots + f^{(C)}_j) \\
  & \geq & (1-\epsilon/r_{K-1}) 2^{K-1-j} x^{(C)}_{K-1} + (2^{K-2-j} f^{(C)}_{K-2}+ \ldots + f^{(C)}_j) \\
  & \geq & (1-\epsilon/r_{K-1}) (2^{K-1-j} x^{(C)}_{K-1} + 2^{K-2-j} f^{(C)}_{K-2}+ \ldots + f^{(C)}_j) \\
  & \ldots & \\
  & \geq & (1-\epsilon/r_{K-1}) (1-\epsilon/r_{K-2}) \ldots (1-\epsilon/r_j) x^{(C)}_j \\
  & \geq & (1-(\epsilon 2^{-j/2})/r_{K-1-j}) (1-(\epsilon 2^{-j/2})/r_{K-2-j}) \ldots (1-(\epsilon 2^{-j/2})/r_0) x^{(C)}_j\quad{\rm since~} r_{j+i} = r_i \cdot 2^{j/2}\\
  & \geq & (1-(\epsilon/3)2^{-j/2}) x^{(C)}_j\quad \rm{by~Fact~\ref{fact:product}.}
\end{eqnarray*}
\end{proof}
\noindent
Theorem~\ref{thm:stage1} now follows as a corollary of the above lemma
for $j=0$.

\noindent
\paragraph{Proof of Theorem~\ref{thm:stage2}.}
Now, we use the sparsification framework developed 
in~\cite{HariharanP10} and outlined previously in 
section~\ref{sec:framework} to prove Theorem~\ref{thm:stage2}. 
Observe that edges $F_0\cup 2^K Y_K$ are identical in 
$G_S$ and $G_{\epsilon}$. Therefore, we do not consider
these edges in the analysis below.

For any $i\geq 1$, let $\psi(i)$ be such that 
$2^{\psi(i)}\leq \rho\cdot 4^i\leq 2^{\psi(i)+1} - 1$. 
Note that for any $j$, $\psi(i) = j$ for at most one
value of $i$. Then, for any $j\geq 1$, $R_j = F_i$
if $j = \psi(i)$ and $R_j = \emptyset$ if there is no
$i$ such that $j = \psi(i)$.
We set $\alpha = 32/3$; $\pi = \rho\cdot 4^K$; for any
$j\geq 1$, $Q_j = (V, W_j)$ where
$W_j = \cup_{i-1\leq r\leq K} 4^{K-r+1}2^r F_r$ if
$R_j \not= \emptyset$ and 
$j = \psi(i)$, and $W_j = \emptyset$ if $R_j = \emptyset$.

The following lemma proves $\pi$-connectivity.
\begin{lemma}
\label{lma:heavy2}
With probability at least $1-4/n$, every edge $e\in F_i = R_{\psi(i)}$
for each $i\geq 1$ is 
$\rho\cdot 4^K$-heavy in $Q_{\psi(i)}$. 
\end{lemma}
\begin{proof}
Consider any edge $e\in F_i$.
Since $F_i\subseteq Y_{i-1}$, Lemma~\ref{lma:heavy} 
ensures that $e$ is $\rho\cdot 2^i$-heavy
in $G_{i-1} = (V, X_{i-1})$, and therefore $\rho\cdot 2^{2i-1}$-heavy in 
$(V, 2^{i-1} X_{i-1})$. Since $\epsilon\leq 1$, Lemma~\ref{lma:sampling1} 
ensures that with probability at least $1-4/n$, the weight of each
cut in $(V, 2^{i-1} X_{i-1})$ is preserved up to a factor of 2 in 
$Z_i = (V, \cup_{i-1\leq r\leq K} 2^r F_r)$. Thus, $e$ is 
$\rho\cdot 4^{i-1}$-heavy in $Z_i$. 

Consider any cut $C$ containing $e\in F_i$. We need to show that the
weight of this cut in $Q_{\psi(i)}$ is at least $4^K$. 
Let the maximum $\lambda_a$ of an edge $a$ in $C$
be $\rho\cdot 4^{k_C}$, for some $k_C\geq i$. By the above proof, $a$ is
$\rho\cdot 4^{k_C-1}$-heavy in $Z_{k_C}$. Then, the total weight of
edges crossing cut $C$ in $Q_{\psi(k_C)}$ is at least 
$\rho\cdot 4^{k_C-1}\cdot 4^{K-k_C+1} = \rho\cdot 4^K$. Since $k_c\geq i$,
$\psi(k_C)\geq \psi(i)$ and $Q_{\psi(k_C)}$ is a subgraph of $Q_{\psi(i)}$. 
Therefore, the the total weight of edges crossing cut $C$ in $Q_{\psi(i)}$ 
is at least $\rho\cdot 4^K$.
\end{proof}

We now prove the $\alpha$-overlap property. For any cut $C$, 
let $f^{(C)}_i$ and $w^{(C)}_i$ 
respectively denote the total weight
of edges crossing cut $C$ in $F_i$ and $W_{\psi(i)}$ 
respectively for any $i\geq 0$. 
Further, let the number of edges crossing
cut $C$ in $\cup_{i=0}^K 2^i F_i$ be $f^{(C)}$. Then, 
\begin{eqnarray*}
\sum_{i=1}^K \frac{w^{(C)}_i 2^{\psi(i)-1}}{\pi} 
& \leq & \quad \sum_{i=1}^K \frac{w^{(C)}_i \rho\cdot 4^i}{2\rho\cdot 4^K} 
\quad = \quad \sum_{i=1}^K \frac{w^{(C)}_i}{2\cdot 4^{K-i}} 
\quad = \quad \sum_{i=1}^K \frac{w^{(C)}_i}{2\cdot 4^{K-i}} 
\quad = \quad \sum_{i=1}^K\sum_{r=i-1}^K \frac{f^{(C)}_r\cdot 2^r\cdot 4^{K-r+1}}{2\cdot 4^{K-i}} \\
= \quad \quad \sum_{i=1}^K \sum_{r=i-1}^K \frac{f^{(C)}_r}{2^{r-2i-1}} 
& = & \sum_{r=0}^K \sum_{i=1}^{r+1} \frac{f^{(C)}_r}{2^{r-2i-1}} 
\quad = \quad \sum_{r=0}^K \frac{f^{(C)}_r}{2^r} \sum_{i=1}^{r+1} 2^{2i+1} 
\quad = \quad \frac{32}{3}\sum_{r=0}^K 2^r f^{(C)}_r 
\quad = \quad \frac{32}{3} f^{(C)}.
\end{eqnarray*}
\noindent
Using Theorem~\ref{thm:framework}, we conclude the proof of 
Theorem~\ref{thm:stage2}.

\noindent
\paragraph{Size of the skeleton graph.}
We now prove that the expected number of edges in $G_{\epsilon}$
is $O(n\log n/\epsilon^2)$. For $i\geq 1$,
define $D_i$ to be the set of connected 
components in the graph $G_i = (V, X_i)$; let $D_0$ be 
the single connected component in $G$. For any $i\geq 1$, if any 
connected component in $D_i$ remains intact in $D_{i+1}$, then there 
is no edge from that connected component in $F_i$. On the other hand, 
if a component in $D_i$ splits into $\eta$ components in $D_{i+1}$, then 
the algorithm explicitly ensures that 
$\sum_{e\in F_i} \frac{w_e}{\lambda_e}$
from that connected component is 
$\sum_{e\in F_i} \frac{2^i}{\rho\cdot 4^i}\leq \left(\frac{\rho\cdot 2^{i+2}\cdot 2^i}{\rho\cdot 4^i}\right)\eta = 4\eta\leq 8(\eta -1)$. Therefore, if $d_i = |D_i|$, then 
\begin{equation*}
\sum_{i=1}^K \sum_{e\in F_i} \frac{w_e}{\lambda_e} 
\leq \sum_{i=1}^K 8(d_{i+1} - d_i) \leq 8n,
\end{equation*}
since we can have at most $n$ singleton components. It follows from
Theorem~\ref{thm:framework} that 
the expected number of edges added to $G_{\epsilon}$ by the sampling
is $O(n\log n/\epsilon^2)$.
Since the number of edges added to $G_{\epsilon}$ in steps~\ref{add1}
and~\ref{loop1} of the algorithm is $O(n\log n/\epsilon^2)$, the 
total number of edges in $G_{\epsilon}$ is $O(n\log n/\epsilon^2)$.

\noindent
\paragraph{Time complexity of the algorithm.}
If $m \leq 2 \rho n$, the algorithm terminates after the first 
step which takes $O(m)$ time. Otherwise, we prove that the expected 
running time of the algorithm is $O(m + n\log n/\epsilon^2) = O(m)$ since 
$\rho = \theta(\log n/\epsilon^2)$. 
First, observe that phase 1 takes $O(m+n\log n)$ time.
We will show that iteration $i$ of phase 2 takes $O(|Y_{i-1}|)$ time.
Since $Y_i\subset X_i$ and $\ex[|X_i|] = \ex[|X_{i-1}|]/2$,
and $|Y_0|\leq m$, it follows that 
the expected overall time complexity of phase 2
is $O(m)$. Finally, the time complexity of phase 3 is 
$O(m+n\log n/\epsilon^2)$ (see e.g.~\cite{KachitvichyanukulS88}).

In iteration $i$ of phase 2, 
the first step takes $|Y_{i-1}|$ time. We show that
all the remaining steps take $O(|X_i| + n\log n)$ time.
Since $X_i\subseteq Y_{i-1}$ and the steps are executed only if
$Y_{i-1} = \Omega(n\log n/\epsilon^2)$, it follows that the total
time complexity of iteration $i$ of phase 2 is $O(|Y_{i-1}|)$.

First, observe that step~\ref{loop1-next} 
and the if block of step~\ref{loop2} take $O(|X_i|)$ time. So, we are 
left with the repeated invocations of the else block of
step~\ref{loop2}.
Each iteration of the else block takes $O(|V_c|\log n+|E_c|)$ time
for the current $V_c, E_c$. So, the last invocation of the else block 
takes at most $O(|X_i| + n\log n)$ time. In any other invocation,
$|E_c| = \Omega(|V_c|\log n)$ and hence the time spent is $O(|E_c|)$.
We show that $|E_c|$ decreases by a factor of 2 from one invocation of 
the else block to the next; then the total time over all invocations 
of the else block is $O(|X_i| + n\log n)$.

To see that the $|E_c|$ halves from one invocation of the else block to 
the next, consider an iteration that begins with 
$|E_c| > 2k_i \cdot|V_c|$. By Lemma~\ref{lma:ni-con},
$E_c$ for the next iteration 
(denoted by $E'_c$) comprises only edges in the first $k_i$ NI forests 
constructed in the current iteration. So 
$|E'_c|\leq k_i\cdot |V_c| < |E_c|/2$.

%% file: future.tex
\section{Future Work}
\label{sec:future}

The obvious open question is whether these results can
be extended to weighted graphs, at least if the weights are
polynomially bounded in $n$. Another possibility is to
extend these results to the semi-streaming model for
unweighted graphs. A more ambitious open problem
is to obtain an efficient (i.e. near-linear in $m$)
algorithm that constructs a skeleton containing
$o(n\log n)$ edges while approximately preserving the weights 
of all cuts with high probability.

%% file: picon.tex
\section{Proof of Lemma~\ref{lma:picon-hp}}
\label{sec:picon}

To prove Lemma~\ref{lma:picon-hp}, we will need two theorems
from~\cite{HariharanP10}. The first theorem is a non-uniform
extension of Chernoff bounds.~\footnote{For Chernoff bounds,
see e.g.~\cite{MotwaniR97}.}

\begin{theorem}[Hariharan-Panigrahi\cite{HariharanP10}]
\label{thm:compression}
Consider any subset $C$ of unweighted edges, where 
each edge $e\in C$ is sampled independently with probability $p_e$
for some $p_e\in [0, 1]$ and given weight $1/p_e$ if selected
in the sample. 
Let the random variable $X_e$ denote the weight of edge $e$ in the
sample; if $e$ is not selected in the sample, then $X_e = 0$. 
Then, for any $p$ such that $p \leq p_e$ for all edges $e$, 
any $\epsilon\in (0, 1]$, and any $N\geq |C|$, the following
bound holds:\footnote{For any event $\cal E$, $\pr[{\cal E}]$
represents the probability of event $\cal E$.}
\begin{equation*}
\pr\left[|\sum_i X_e - |C|| > \epsilon N\right] < 2 e^{-0.38 \epsilon^2 pN}.\qedhere
\end{equation*}
\end{theorem}
\noindent
To state the second theorem, we need the following definitions.
\begin{definition}
For any undirected graph $G$ and 
for any $k > 0$, the $k$-projection of any cut $C$ is the set
of $k$-heavy edges in $C$.
\end{definition}
\begin{definition}
The {\em edge connectivity} of an undirected graph $G$ is the minimum weight 
of a cut in $G$.
\end{definition}
\noindent
The theorem counts the number of distinct $k$-projections in cuts of weight
$\alpha k$ for any $k\geq c$, where $c$ is the edge connectivity of
the graph.
\begin{theorem}[Hariharan-Panigrahi\cite{HariharanP10}]
\label{thm:cut-count-new}
For any undirected graph with edge connectivity $c$ and for any 
$k \geq c$ and any $\alpha \geq 1$, the 
number of distinct $k$-projections of cuts of weight at most 
$\alpha k$ is at most $n^{2\alpha}$.
\end{theorem}
\noindent
Using the above two theorems, we now prove Lemma~\ref{lma:picon-hp}.

\begin{proof}[Proof of Lemma~\ref{lma:picon-hp}]
Let ${\cal C}_j$ be the set of all cuts $C$ such that 
$2^j \pi \leq r^{(C)} \leq 2^{j+1}\pi-1$, $j\geq 0$. 
We will prove that with probability at least $1 - 2n^{-2^{j+1}}$,
all cuts in ${\cal C}_j$ satisfy the property of the lemma.
Then, the lemma follows by using the union bound over $j$ since 
$2n^{-2} + 2n^{-4} + \ldots + 2n^{-2j} + \ldots \leq 4n^{-2}$.

We now prove the property of the lemma for cuts $C\in {\cal C}_j$. 
Since each edge $e\in R^{(C)}$ is sampled with probability
$p$ in obtaining $\widehat{R^{(C)}}$, 
we can use Theorem~\ref{thm:compression}
with sampling probability $p$. Then, for any
$R^{(C)}$ where $C\in {\cal C}_j$, by Theorem~\ref{thm:compression}, 
we have
\begin{equation*}
\pr\left[\left|\widehat{r^{(C)}} - r^{(C)}\right| > \delta q^{(C)}\right] <
2e^{-0.38\cdot\delta^2\cdot p \cdot q^{(C)}} \leq
2e^{-0.38\cdot\delta^2\cdot p \cdot \pi\cdot 2^j} \leq
2e^{-6\cdot 2^j\ln n} =
2n^{-6\cdot 2^j},
\end{equation*}
since $q^{(C)}\geq \pi\cdot 2^j$ for any $C\in {\cal C}_j$.
Since each edge in $R^{(C)}$ is 
$\pi$-heavy in $(V, Q)$,  
Theorem~\ref{thm:cut-count-new} ensures that 
the number of distinct $R^{(C)}$
sets for cuts $C\in {\cal C}_j$ is at most 
$n^{2\left(\frac{\pi\cdot 2^{j+1}}{\pi}\right)} = n^{4\cdot 2^j}$.
Using the union bound over these distinct $R^{(C)}$ edge sets,
we conclude that with probability at least $1 - 2n^{-2^{j+1}}$,
all cuts in ${\cal C}_j$ satisfy the property of the lemma.
\end{proof}